# Ultra-high-precision fused silica micro-hole machining via spherical aberration-assisted filamentation and laser-induced deep etching

Seunghyun Bang[1], Seonghyeon Kang[1], Hyunjong Lee[1], Hyungsik Kim[2], Seokho Song[1], Kwang-Geol Lee[1,*]

[1]Department of Physics, Hanyang University, Seoul 04763, Republic of Korea
[2]Samsung Display Co., Ltd., Giheung-gu, Yongin-si, Gyeonggi-do 17113, Republic of Korea

*kglee@hanyang.ac.kr

## Abstract

Glass materials play an increasingly important role in advanced technologies due to their superior physical properties. However, precise machining of glass remains a major challenge because of its brittleness and sensitivity to thermal and mechanical stresses. In this study, we present a novel approach that combines spherical-aberration–assisted filamentation with Laser-Induced Deep Etching (LIDE) to achieve unprecedented high-precision micro-hole machining in fused silica substrates. By deliberately introducing spherical aberration into an intense femtosecond laser beam, thin, uniformly elongated, and stable filaments are generated, which effectively suppress unwanted plasma formation and thermal deformation typical of standard filamentation. Using this method, we fabricated micro-holes with diameters as small as 10 μm across various sizes, maintaining an almost zero taper even in 1 mm-thick samples. The sidewalls exhibited nanoscale smoothness ($R_a$ = 38.1 nm, RMS = 53.9 nm), and the hole area demonstrated excellent repeatability with only ~1.0% variation across multiple trials. This simple optical configuration drastically reduces cost compared with existing approaches that rely on specialized components, while moderately satisfying critical requirements for geometrical versatility, minimal damage, precision, and repeatability. This work represents a significant step forward in precision glass machining and lays a foundation for future microstructured electronic, optical, and microfluidic devices.

## Introduction

Recently, glass materials have garnered renewed attention as essential components in various advanced technological fields, including electronics, optics, and biotechnology [1,2]. This resurgence is attributed to the unique physical properties of glass—such as excellent electrical insulation, high optical transparency, chemical and thermal stability, and mechanical strength—that facilitate the advancement and high integration of next-generation devices [3]. These characteristics position glass as a promising alternative capable of overcoming limitations inherent to silicon-based materials in numerous applications, including semiconductor packaging, high-resolution displays, and highly sensitive sensors [4]. In semiconductor applications, specifically, the use of through-glass vias (TGV) enables the realization of three-dimensional integrated circuits (3D IC) and high-density packaging technologies [5]. In display technologies, glass substrates facilitate micro-apertures and micro-lens arrays, optimizing resolution and luminance for OLED and micro-LED displays [6]. Additionally, in sensor technologies, glass-based microstructures increase reactive surface area, thereby enhancing sensitivity and selectivity [7].

Despite these advantages, practical utilization of glass materials has encountered fundamental technical limitations. Due to its amorphous nature, glass exhibits brittleness and is highly susceptible to cracking or fracturing during mechanical processing, making high-precision machining challenging [8]. Micro-hole machining for applications in optics, microfluidics, electronic packaging, and especially for through-glass vias (TGVs) in 3D IC integration must simultaneously satisfy precise control over hole geometry and size [9], minimal damage to the surrounding glass [10], near-vertical taper angles [5], ultra-smooth sidewalls [11,12], and high repeatability [9,13]. Conventional processing techniques, including laser ablation, dry etching, and wet etching, have demonstrated clear limitations in achieving the resolution and process controllability required for fabricating micro- and nanoscale structures [14-16]. These techniques have yet to mature sufficiently to simultaneously satisfy diverse requirements, including precision, repeatability, minimized thermal impact, and processing speed [17].

In this context, various laser-based glass processing techniques have been proposed. Among these, Laser-Induced Deep Etching (LIDE)—which involves selective internal modification of glass by irradiating high intense laser beam followed by chemical etching—has received considerable attention for its high precision [17]. However, conventional LIDE methods utilizing focused Gaussian beams necessitate spot-by-spot scanning, resulting in significant processing times even for relatively simple structures, thus limiting efficiency [18]. To overcome these limitations, high-speed glass hole-drilling methods employing Bessel beams [19,20] and laser filamentation have been developed [21-23]. Although these approaches greatly reduce processing time for simple hole structures, Bessel beam methods require specialized optical components, such as spatial light modulators (SLMs) [24,25], axicon lens–based Bessel beams [26], or diffractive optical elements (DOEs) [27], leading to reduced stability and limited control [28]. Furthermore, existing filamentation techniques suffer from non-uniform energy distributions and thermal deformation caused by plasma formation, impeding precise shape control [29].

Therefore, innovative technological approaches capable of concurrently ensuring precision, speed, and controllability in glass processing are required. In this study, we propose a novel high-precision machining method that intentionally introduces spherical aberration to control laser filamentation, subsequently integrating it with the LIDE process. By utilizing spherical aberration, we can effectively modulate the phase distribution of the laser beam to form thin, uniform, and stable filaments extending over the millimeter scale before the focal region. This significantly mitigates issues related to irregular plasma formation and thermal deformation observed in conventional filamentation techniques, thereby enabling unprecedented precision in micro-hole machining. Such a filament with micrometer-scale width can be extended over several millimeters, resembling a slender light blade, similar to a miniaturized Jedi sword in Star Wars. Hereafter, we refer to this as the 'pre-filament'.

Both theoretically and experimentally, we investigate how the presence or absence of spherical aberration affects filament length and morphology, and demonstrate enhanced machining performance when these filaments are applied to the LIDE process. Ultimately, this research aims to establish high-precision, high-efficiency glass machining technology, thereby contributing to the advancement of microstructure-based electronic and optical devices.

## Results

### *Performance of our protocol*

Figure 1 illustrates our two-step process. In the first step (Fig. 1a), a 1030 nm femtosecond laser was focused through a lens with designed spherical aberration. A fused silica substrate (1 mm thick) was positioned in the pre-filament zone, and circular patterns corresponding to the target hole diameters were inscribed by raster-scanning the substrate in the x–y plane. In the second step, the laser irradiated glass was immersed into a wet etching solution to selectively remove the laser-modified regions, thereby releasing the filament traces as through-holes. As shown in Fig. 1c, when the glass is etched along the filament traces, it separates into a plate with through-holes and glass rods freed from those holes. During the cleaning process, these rods are flushed out and removed during the cleaning process.

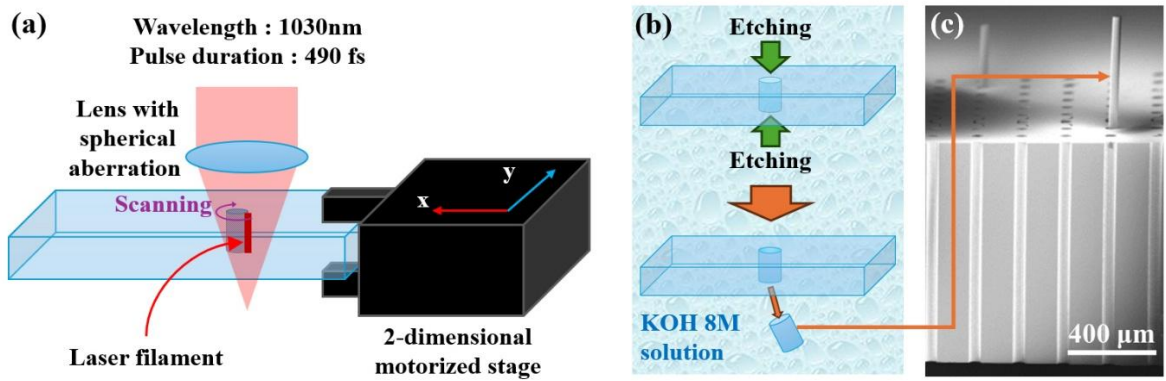

**Figure 1** High-precision micro-hole fabrication in fused silica using aberration-assisted Laser-Induced Deep Etching (LIDE): **(a)** Laser Irradiation. A femtosecond laser (1030 nm) is focused into a fused silica substrate. By intentionally utilizing the spherical aberration of a lens system, a thin and stable laser filament is generated within the glass. The substrate is then raster-scanned using a 2D stage to inscribe a pre-defined pattern, creating a localized internal modification. **(b)** Selective Etching. The laser-irradiated substrate is immersed in an 8 M potassium hydroxide (KOH) solution. The laser-modified region exhibits a significantly higher etch rate, leading to its selective removal and the formation of through-holes as the inner glass rod is separated. **(c)** Fabrication Result. A cross-sectional scanning electron microscope (SEM) image captures the clean separation of the glass rods during the etching process. This demonstrates the formation of through-holes with highly vertical sidewalls and minimal damage, highlighting the precision of the technique.

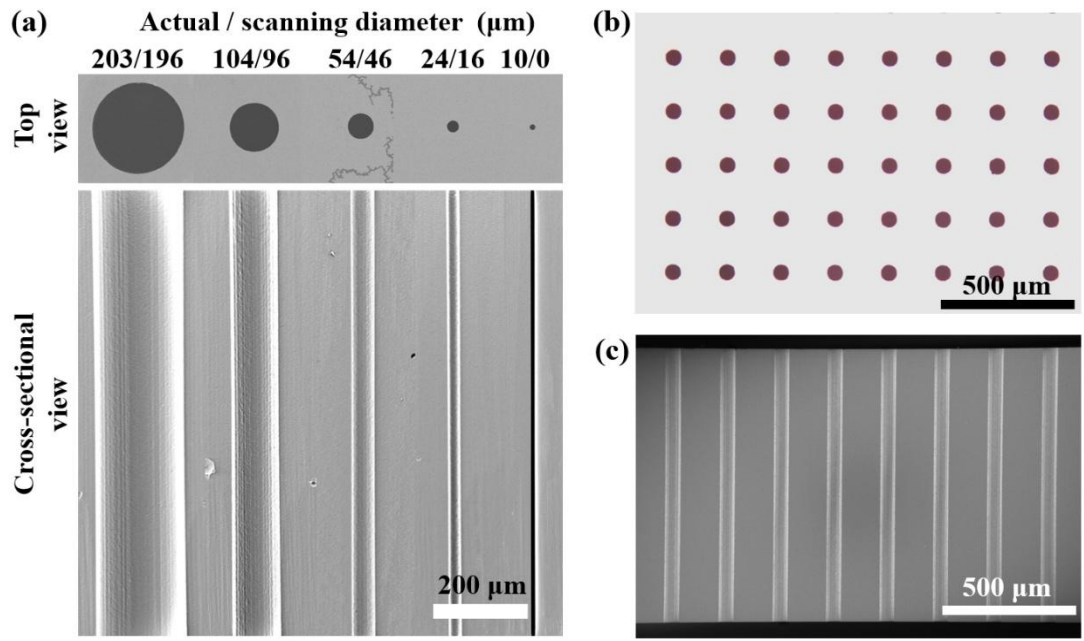

**Figure 2**. Geometrical precision and repeatability of micro-holes fabricated by aberration-assisted LIDE: **(a)** Top-view and cross-sectional SEM images of fabricated micro-holes with final diameters of 10, 24, 54, 104, and 203 μm. These were created from scanning diameters of 0, 16, 46, 96, and 196 μm, respectively; the ~8-10 μm increase in diameter results from the wet etching step. The top-views confirm crack-free processing, while the cross-sections demonstrate highly uniform profiles with near-zero taper angles. **(b)** Repeatability assessment using a 5 × 8 array of 54 μm diameter holes. Analysis of 160 holes fabricated under identical conditions revealed a hole area variation of only 1.00% (standard deviation), confirming excellent process stability. **(c)** Taper angle consistency test. The cross-sectional profiles of multiple holes show nearly perfectly vertical sidewalls, with a taper angle below the SEM's resolution limit (effectively ≲ 0.1°), demonstrating high uniformity across the array.

Figure 2a shows the profiles of fabricated circular holes with a wide range of diameters from a minimum of 10 μm up to 200 μm, demonstrating the capability to produce holes of various shapes and sizes. The smallest hole diameter of 10 μm reflects the resolution limit for scanning-based patterning. This lower bound, optimized within the trials conducted in this study, can be further improved by adjusting parameters such as the laser wavelength and pulse width, the numerical aperture (NA) of the lens, the longitudinal spherical aberration (LSA), and the type and concentration of the etching solution [20]. Importantly, no cracks or fractures were observed around any hole, confirming that the machining process caused no collateral damage. This indicates that the structural integrity of the surrounding glass was preserved, ensuring the reliability and durability of any subsequent device. In the cross-sectional SEM images (Fig. 2a), the hole diameter remained uniform throughout the entire depth, and any taper was too small to be resolved within the SEM's resolution. Based on the pixel size (1.5 μm/pixel) and the observed hole depth, this corresponds to an angular detection limit of ≈0.08°, which confirms that the taper angle lies below the ideal criterion of ≲0.1°. To assess repeatability, we patterned 54 μm holes in four 5 × 8 arrays (160 holes in total, not shown) under identical conditions, and all of which were formed reliably as shown in Fig. 2b. The standard deviation of the hole area was 21.9 μm² (~1.0 %), demonstrating excellent process stability. Furthermore, to verify taper-angle repeatability, we measured ten holes and found their sidewalls to be almost perfectly vertical, below the SEM's resolution limit, demonstrating highly consistent results (Fig. 2c).

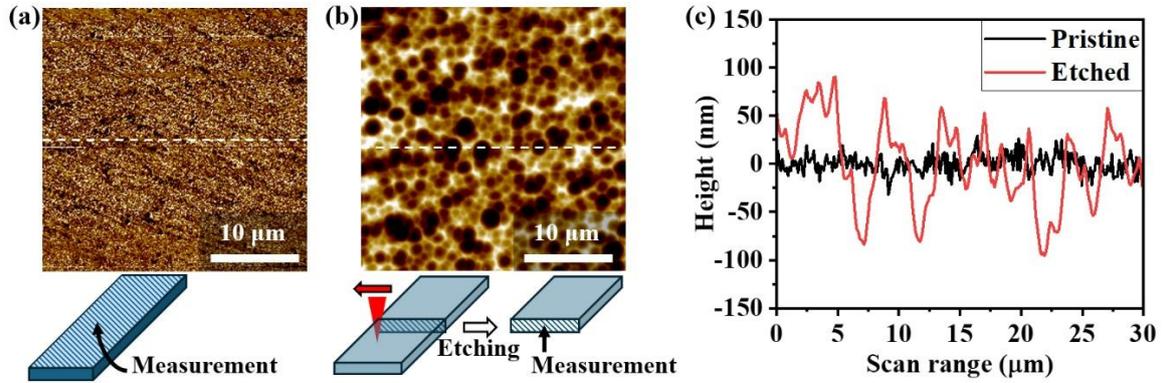

**Figure 3** AFM analysis comparing the surface topography of pristine and processed glass. AFM images (30 × 30 μm²) of a pristine, unprocessed fused silica surface **(a),** and of an etched sidewall of a fabricated micro-hole **(b)**, showing that the surface remains exceptionally smooth even after processing. **(c)** Corresponding surface height profiles measured along the dashed lines in (a) and (b). The profile of the etched sidewall (red line) shows slightly increased roughness compared to the pristine surface (black line) but maintains nanoscale smoothness.

The sidewall surface roughness was quantitatively measured over a 30 × 30 μm² area using atomic force microscopy (AFM) (Fig. 3). The initial glass surface before processing had a roughness of Ra = 9.07 nm and RMS = 11.9 nm, which is a value similar to that of commercial slide glass. In contrast, the sidewall formed through the etching process was measured to have a roughness of $R_a$ = 38.1 nm and RMS = 53.1 nm (Table 1). Although these values fall short of the few-nanometer requirement for ultra-precision optics and photonics, they nonetheless demonstrate surface quality sufficient for practical applications such as glass interposers in semiconductor packaging and microfluidic devices.

| (Unit: nm) | $R_a$ | RMS |
|---|---|---|
| **Pristine** | 9.07 | 11.9 |
| **Etched** | 38.1 | 53.9 |

**Table 1** The measured average ($R_a$) and root-mean-square (RMS) roughness values for the pristine and etched surfaces. The final sidewall roughness ($R_a$ = 38.1 nm, RMS = 53.9 nm) meets the requirements for applications such as glass interposers and microfluidic.

*Effect of optical aberrations on filamentation characteristics*

Laser filamentation is a nonlinear optical phenomenon that occurs when a laser beam with peak power above the critical power propagates through a medium [30]. Filamentation arises from the interplay of several nonlinear effects primarily self-focusing due to the Kerr effect, defocusing caused by the high-intensity generated plasma, and multiphoton absorption (MPA) [31]. When these competing effects reach a dynamic balance, the beam undergoes repeated cycles of self-focusing and defocusing, forming a stable plasma channel during propagation [32]. This process can be described mathematically as follows [33]:

$$\frac{\partial A}{\partial z} = \frac{i}{2k_0}\nabla_\perp^2 A - \frac{\alpha}{2}A - \frac{i\beta_2}{2}\frac{\partial^2 A}{\partial t^2} + ik_0 n_2 |A|^2 A - \frac{k_0 n_2}{\omega_0}\frac{\partial(|A|^2 A)}{\partial t} + ik_0 \Delta n_{plasma} A - \frac{\beta_K}{2}|A|^{2K-2}A \quad (1)$$

Here, $A$ is the slowly varying complex envelope of the electric field, $z$ is the propagation coordinate, and $k_0$ and $\omega_0$ are the vacuum wavenumber and central angular frequency, respectively. The coefficient $\alpha$ denotes linear absorption, $\beta_2$ represents group velocity dispersion, $n_2$ is the Kerr nonlinear index, $\beta_K$ is the $K$-photon absorption coefficient, and $\Delta n_{plasma}$ is the refractive-index change induced by free electrons under the Drude model.

When focusing a beam with an aberration-corrected lens, filamentation begins at the point of maximum energy density during propagation, or the beam collapses due to the resulting excessive plasma density [34]. If this is applied to material processing, the target material is necessarily positioned within this maximum energy density zone, making it difficult to avoid unnecessary over-processing or material damage. To mitigate this issue, research has been conducted on applying a separate mask on the material to induce filamentation [35].

In this study, we address this issue by intentionally introducing optical aberrations. Specifically, we induce an LSA, to generate a filament before the focal point, enabling finer and more uniform processing [35]. This approach removes the need for the target material to be located at the point of maximum energy density. To investigate and validate this phenomenon, we conducted a comparative study through simulations and experiments using an aberration-corrected lens and a spherical-aberrated lens with similar numerical apertures (NA).

In this simulation, we solved the nonlinear Schrödinger equation, shown in Eq.(1), in the radial–axial (r–z) space using the Axisymmetric Split-Step Hankel Method (ASSHM). This method consists of three steps at each propagation stage: (1) applying a phase kick in the spatiotemporal domain that accounts for virtually all nonlinear effects, (2) efficiently performing radial diffraction using the Hankel transform, and then (3) reapplying a phase kick. The propagation medium was modeled with the optical properties of fused silica. Compared to conventional four-dimensional (3D + t) simulations, the ASSHM offers the advantage of quantitatively predicting filamentation phenomena with axisymmetric structures while significantly reducing the computational load. (For details of the algorithm, see Supplementary Information S1.1)

The initial field was modeled as an ideal Gaussian electric field with added focusing and aberration terms. A temporal term was also included to account for group-velocity dispersion (GVD), plasma formation, and multiphoton absorption. The wavefront aberration function and Zernike coefficients for the aberrated lens were calculated using Zemax. Due to the two-dimensional nature of the simulation, which cannot represent asymmetric aberrations, only cylindrically symmetric aberrations were included. Furthermore, amplitude and phase noise were introduced into the Gaussian beam. The resulting expression for the initial field is given as follows:

$$A(r,t) = \sqrt{I_0}e^{-\frac{r^2}{w_0^2}} \cdot e^{-i\frac{k_0 r^2}{2f}} \cdot e^{-i2\pi\Phi_{abr}} \cdot \mathrm{sech}\left(\frac{t}{t_0}\right) \cdot [1+\delta(r)]\, e^{i\varphi(r)} \quad (2)$$

In this equation, $I_0$ is the peak intensity, $w_0$ is the beam waist radius at the lens, and

$e^{-i\frac{k_0 r^2}{2f}}$ represents the quadratic phase imparted by a lens of focal length *f*, while $e^{-i2\pi \Phi_{abr}}$ encodes the time-dependent phase shift due to spherical aberration quantified by $\Phi_{abr}$. The factor sech(*t*/*t₀*) term describes the temporal pulse shape with characteristic half-duration $t_0$. Finally, the factor $[1 + \delta(r)]$ and $e^{i\varphi(r)}$ represent the amplitude and phase noise terms, respectively.

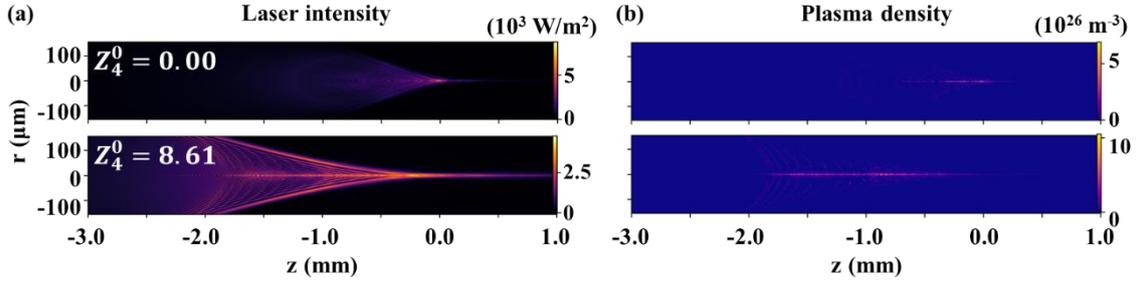

**Figure 4** A numerical simulation was conducted to compare filamentation and plasma formation processes under aberration control, in which a femtosecond laser pulse was focused into fused silica using two different lens systems (NA ≈ 0.25). The simulations were performed using the Axisymmetric Split-Step Hankel Method (ASSHM), with *z*=0 representing the geometric focal position. The simulation results show that in an ideal, aberration-free system ($Z_4^0 = 0.00$), the laser intensity (a) is sharply focused at the geometric focal point, resulting in the localized formation of a short, high-density plasma (b). In contrast, in a system where spherical aberration was intentionally introduced ($Z_4^0 = 8.61$), the laser intensity (a) is reshaped into a long, stable 'pre-filament' well before the focal point, which in turn creates a long, thin plasma channel (b) with a lower and more uniform density.

Figure 4 compares the effect of spherical aberration on the propagation of a femtosecond Gaussian beam inside fused silica (detailed simulation parameters are given in Supplementary Information S1.2). In the ideal, aberration-free case ($Z_4^0 = 0$), the beam energy is focused at the geometrical focal plane (*z*=0) with the formation of a short filament. The filament generated here corresponds to that produced by self-induced nonlinear effects, resulting from the focal shift caused by the Kerr effect. Consequently, a localized, high-intensity plasma forms near around the focal point. As plasma density approaches the critical density, the electric field undergoes strong defocusing and energy absorption due to the rapid change in refractive index and multiphoton absorption by the plasma, causing the propagation to be suppressed and terminated [36]. This is why the plasma channel stops shortly after the focal point. In contrast, when LSA is intentionally introduced ($Z_4^0 = 8.61$), a long and narrow filament forms throughout the aberrated focal region. Specifically, as the Zernike $Z_4^0$ value (the degree of spherical aberration) increased, the length of the laser filament increased. In contrast, the filament's line width remained constant at approximately 2.5 μm for all $Z_4^0$ values, showing no significant change. (For more details, please refer to Supplementary Information S1.3). Following the filament profile, a thin and long plasma channel is generated mostly prior to the focal point (Fig. 4b). Our results suggest that this plasma seems to induce uniform modification inside the glass, enabling the formation of more precise structures.

To achieve uniform processing within this aberrated region, minute initial phase perturbations in the beam play a decisive role. This is because real laser pulses, unlike ideal gaussian pulses, always contain slight wavefront distortions [37]. If a simulation uses perfectly symmetric initial conditions (no phase error), a non-physical "channel" where all energy is concentrated at a single location can be generated. Therefore, to emulate experimental conditions, this simulation intentionally introduces phase noise to the input beam, which serves as a seed for nonlinear effects such as spatial modulation instability. While this noise induces random inhomogeneities in individual filaments, averaging the results over multiple pulses improved the overall uniformity. All simulation results shown represent the cumulative average of 10 pulses. The standard deviation of the phase noise used was approximately $10^{-3}$ rad, corresponding to a wavefront error of $\lambda/1000$. Although this is significantly better than the typical $\lambda/10$ quality of common experimental optics [38], it is large enough to form significant macroscopic structures after the nonlinear amplification process.

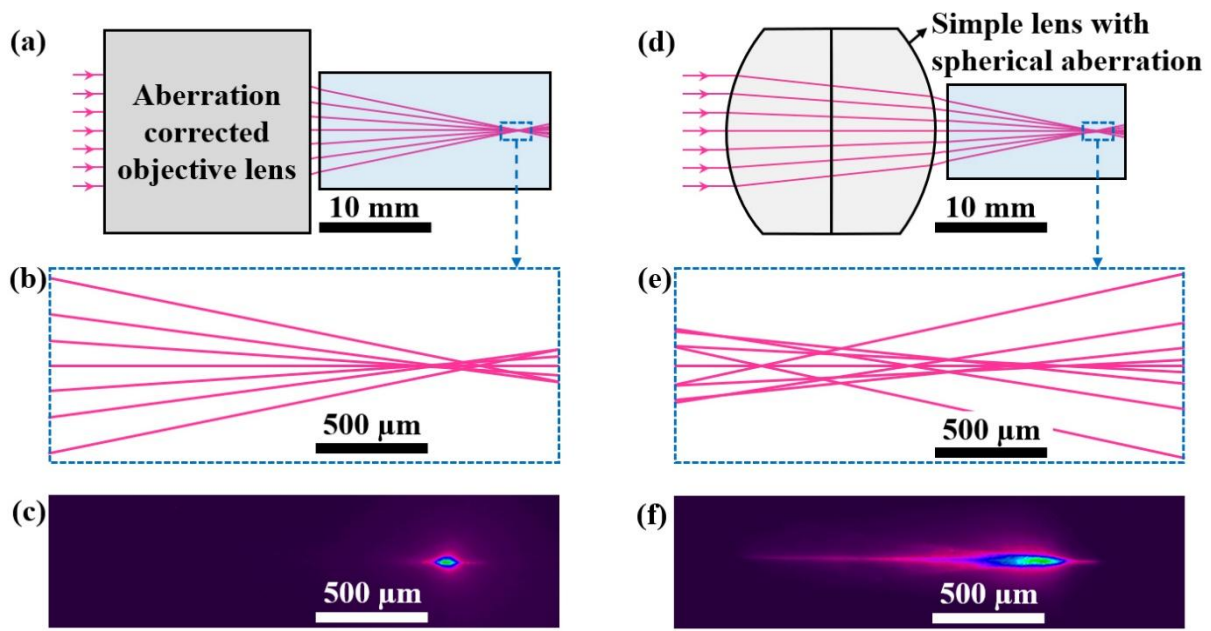

**Figure 5** An experimental comparison of laser filament profiles from two systems (NA ≈ 0.25) validates the simulation trends for aberration-assisted filamentation shown in Figure 4. **(a, b, c)** Aberration-corrected system: **(a)** Schematic of an aberration-corrected objective lens. **(b)** The corresponding Zemax ray-tracing simulation shows that light rays converge to a single, ideal focal point. **(c)** The resulting experimental filament, imaged in a fluorescent medium, is short and localized, consistent with the simulation. **(d, e, f)** Spherical aberration-induced system: **(d)** Schematic of the lens system intentionally designed to induce spherical aberration. **(e)** The Zemax simulation reveals a broad longitudinal spherical aberration (LSA) profile, creating an extended focal region. **(f)** The experimentally generated filament is significantly elongated and stable, closely matching the simulation trend and demonstrating effective filament control.

To validate these trends experimentally, we performed measurements with two lenses of nearly identical NA: one aberration-corrected and the other intentionally spherical-aberrated. Fig. 5a,b,d,e show LSA profiles calculated using Zemax for each lens. The aberration-corrected lens exhibits negligible aberration, making filamentation difficult, whereas the spherical-aberration lens provides a broad LSA range that promotes filament formation. The experimental results (Fig. 5c,f) closely match the simulation trends, confirming that assisted spherical aberration plays a key role in inducing and stabilizing laser filamentation. For more details, please refer to the Materials and Methods section.

## Discussion

In this study, we demonstrated a novel approach that integrates spherical-aberration–assisted filamentation with the LIDE process to achieve ultrahigh-precision micro-hole machining in glass. By deliberately introducing spherical aberration into the focusing system, stable and uniform filaments were generated and employed as a machining tool, thereby moderately satisfying the critical requirements of precision shape control, minimal damage, near-vertical taper angle, smooth sidewalls, and high repeatability.

The fabricated micro-holes exhibited taper angles consistently below 0.1° and sidewall roughness levels of $R_a$ = 38.1 nm. These results not only meet the stringent requirements of through-glass vias (TGVs) for 3D IC packaging and high-precision optical components, but also surpass previously reported performances obtained with Gaussian-beam-based LIDE and Bessel-beam drilling methods [5,21-25]. Furthermore, an array of 160 holes demonstrated excellent reproducibility, with only 1.0% variation in hole area and a failure rate of 0%, underscoring the robustness and industrial applicability of this method.

Beyond its practical achievements, the introduced pre-filament functions as a millimeter-long and micrometer-thin optical blade, extending the concept of laser-based processing tools. Through both ASSHM simulations and experiments, we systematically analyzed the formation and stabilization mechanisms of aberration-assisted filaments, deepening the understanding of spherical-aberration–mediated nonlinear optical interactions. This approach provides unprecedented precision and flexibility in glass microdevice fabrication, opening new possibilities across optics, electronic packaging, and microfluidics.

While this study has achieved excellent results using fused silica, several challenges remain for future work. A primary objective is to realize a nanometer-scale surface roughness, which is a critical requirement for optical and photonic applications, by further optimizing the etching process or applying additional planarization treatments. Furthermore, the process must be adapted and optimized for various glass materials with distinct properties—such as borosilicate, alkali-free, and Foturan—to meet the demands of specific applications. Addressing these challenges through subsequent research will enhance processing speed and precision, broaden the range of applications, and ultimately open up new possibilities in the design and fabrication of next-generation glass-based devices.

## Materials and methods

### *Materials*

Fused silica substrates (Thorlabs MS10FS) were used in the experiment. Isopropyl alcohol (98%) and acetone (99.5%) were used for cleaning, and the etchant solution was prepared by dissolving ACS reagent grade potassium hydroxide (≥85%, pellets) to a concentration of 8 M. All chemicals were supplied by Sigma-Aldrich, and deionized (DI) water for cleaning and solution dilution was produced with a resistivity of 18.2 MΩ·cm using a Wellix Plus II system (JEIO Tech).

*Sample preparation*

In this experiment, the condition of the glass surface is crucial for reproducible processing because some particles or grease can cause errors in the experimental results due to burning and scattering when the laser is incident on those areas. Therefore, the glass was first cleaned by rinsing it with deionized (DI) water for 30 seconds to remove dust and other particles. Then, we wiped the glass surface with a cleaning tissue soaked in acetone and isopropyl alcohol sequentially. Then, the sample was further cleansed in an ultrasonic cleaner with DI water for 10 minutes, followed by drying it in a forced convection oven (SH Scientific SH-DO-149FG) at 80 °C for 1 hour.

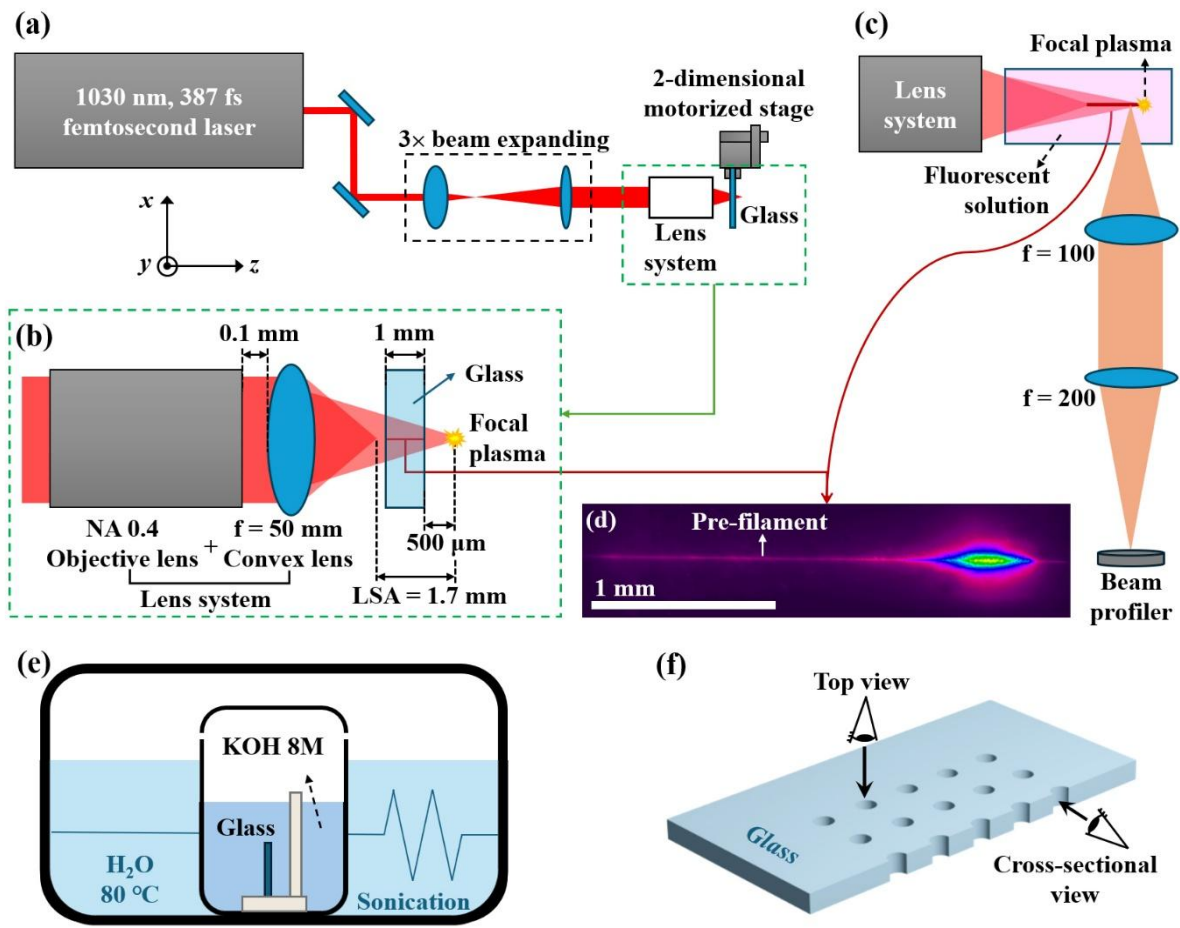

**Figure 6** Schematic of the complete experimental setup and process flow. **(a)** Laser Irradiation System: The overall optical path. A femtosecond fiber laser (Amplitude Tangor 100; wavelength: 1030 nm, pulse duration: 387 fs) is used as the light source. For all experiments, the pulse energy and repetition rate were maintained at 100 μJ and 40 kHz, respectively. **(b)** Detailed Focusing System for Machining: A combination of a 0.4 NA objective lens (Mitutoyo NIR 20×) and a 50 mm plano-convex lens (Thorlabs LA1131-AB) is used to intentionally induce spherical aberration and focus the beam into the fused silica sample. **(c)** Filament Profile Measurement System: The setup for imaging the filament shape (as shown in Fig. 5) consists of a fluorescent solution, a 2x magnification imaging system, and a beam profiler (Ophir SP620U). **(d)** Generated Laser Filament: An image of the actual filament produced by the lens system in (b). The filament has a diameter of approximately 1.8 μm and a length exceeding 2 mm, acting as an optical blade for machining. **(e)** Ultrasonic-Assisted Wet

Etching: The laser-irradiated sample is etched in an 8 M KOH solution at 80 °C for 2 hours. Sonication (37 kHz, 380 W) is applied to facilitate the removal of debris and ensure uniform etching. **(f)** Micro-hole Characterization: To analyze the results, the processed sample is cleaved through the center of the holes, and the cross-sections are coated with gold for SEM imaging.

*Laser system*

Figure 6 shows a schematic of the entire optical setup used in this study. A 1030 nm fiber-amplified femtosecond laser (Amplitude Tangor 100, pulse width: 387 fs) served as the light source. Throughout all experiments presented in this study, the pulse energy was maintained at 100 μJ and the repetition rate at 40 kHz to induce filamentation.

For micro-hole machining, the lens system (Fig. 6b) comprised a 0.4 NA objective lens (Mitutoyo NIR 20×) and a 50 mm focal-length plano-convex lens (Thorlabs LA1131-AB). The two lenses' flat surfaces were separated by 0.1 mm to introduce an LSA of approximately 1.7 mm. Under these conditions, as shown in Fig. 6d, the laser filament formed inside the fluorescent medium measured 1.8 μm in diameter and exceeded 2 mm in length.

As shown in Fig. 6b, the glass substrate was mounted such that its rear surface was located 500 μm before the effective focus (i.e., the point in air at which the plasma intensity reaches its maximum). This configuration prevents unwanted damage to the glass surface from plasma formation and ensures stable filament generation within the bulk material. For hole drilling via the filament, the substrate was affixed to a motorized stage (Precibeo GO XYZ-LS4545) and raster-scanned in the *x–y* plane. In all experiments, the stage scan speed was set to 200 μm/s.

To measure the laser filament propagation profile with and without spherical aberration, two lens systems, as shown in Fig 5, were used. A spherical aberration–corrected objective lens (Thorlabs LMH-10X-1064, NA 0.25) was employed to record the beam propagation profile under near-aberration-free conditions. To intentionally introduce spherical aberration, two plano-convex lenses (f = 30 mm, Thorlabs LA1805-AB) were arranged with their plano surfaces facing each other at a 0.1 mm separation to achieve an effective NA of 0.24. The beam diameter incident on both the aberration-free lens system and the lens system with spherical aberration was adjusted to 10 mm.

For imaging the filament shape, the laser beam was directed into the fluorescent solution using each lens configuration to capture the resulting fluorescence signal. The fluorescent solution was prepared by diluting the fluorophore to 1 % in deionized water and then contained in a quartz reservoir positioned along the beam path to induce filament-generated fluorescence. The propagation profile of the filament within the fluorescent medium was recorded using a two-stage imaging setup (a 100 mm Thorlabs convex lens followed by a 200 mm convex lens) and a beam profiler (Ophir SP620U) configured as a 2× magnification vision system, as illustrated in Fig. 6c. Note that a clear filament image could not be obtained in glass due to burst-induced chaotic features (see Supplementary Information S2).

*Wet etching and post-cleaning*

As shown in Fig. 6e, wet etching was carried out in an 8 M KOH solution at 80 °C for 2 hours. During etching, sonication was applied using an ultrasonic cleaner (Elmasonic P-300H) set to a frequency of 37 kHz and a power of 380 W, with sweep mode enabled to ensure uniform ultrasonic distribution throughout the tank [39]. Teflon beakers and glass holders were used to prevent corrosion and contamination. After etching, the glass substrates were first rinsed with deionized (DI) water. Residual KOH was then removed by sequential rinses in acetone and isopropyl alcohol. Following a final DI water rinse to eliminate any remaining contaminants, the substrates were dried in a forced-convection oven (SH Scientific SH-DO-149FG) at 80 °C for 20 minutes.

*Characterization of micro-holes*

For the SEM images shown in Fig. 2, the fabricated samples were cleaved through the center of the holes as shown in Fig. 6f, and the resulting faces were coated with a 20 nm layer of gold (Au). Low-voltage, high-resolution scanning electron microscopy (SEM, Carl Zeiss Sigma 300) was then used to acquire top-view and cross-sectional images. Top-view images were analyzed to determine hole diameters and to check for any cracks or deformations in the vicinity of each hole. Cross-sectional images were used to measure the taper angle of the holes and also to inspect for any cracks or deformations along the sidewalls.

To evaluate the repeatability of the taper angle, cross-sectional images of 10 different holes were analyzed. Using Python, the boundaries of the hole sidewalls were extracted from each image, and the distances between the opposing sidewalls were measured at various depths to observe the expected diameter reduction toward the center of the hole. However, due to the resolution limit of the SEM images (1280 × 720), no significant variation in distance could be detected.

Sidewall surface roughness was measured using an atomic force microscope (AFM, Park Systems XE-100) operating in non-contact mode over a 30 × 30 μm$^2$ area. Both a pristine (unetched) glass surface and an etched sidewall region, as illustrated in Fig. 3, were scanned under identical conditions, and the root-mean-square (RMS) roughness and arithmetic mean roughness ($R_a$) were computed for comparison.

Hole diameter repeatability was assessed using an optical microscope (Nikon LV150N). A top-view image of the hole grid was captured at 10× magnification. Python was used to generate a mask for each hole, and the area enclosed by each mask was calculated. A total of 160 holes were measured, and the resulting area distribution was used to quantify the repeatability error.

**Acknowledgements**

This work was supported by Creation of the Quantum Information Science R&D Ecosystem (Grant No. RS-2023-NR068116) through the National Research Foundation of Korea (NRF) funded by the Korean government (Ministry of Science and ICT), and by the Institute of Information and Communications Technology Planning & Evaluation (IITP) grant funded by the Korean government (MSIT) (RS-2022-II221026, RS-2025-02215576). The authors also gratefully acknowledge Samsung Display Co., Ltd. for providing laser instrumentation used in this research.

**Author Contributions**

S.B. and K.-G.L. conceived the concept and initiated the project. S.B. and H.K. designed and built the laser system for the experiments. S.B. and S.K. prepared the glass samples, conducted all experiments and measurements, and analyzed the acquired data with K.-G.L. S.B. and H.L. developed the nonlinear optical propagation simulation code and acquired the simulation data. S.B., S.S., and K.-G.L. proposed the theoretical model for laser filamentation and led the discussions. All authors contributed to the writing of the manuscript, and K.-G.L. supervised the overall project.

**Conflict of interest**

The authors declare no conflict of interest.

**Data availability**

The main data supporting the results of this study are available within the paper and its Supplementary Information. Other raw data generated during this study are available from the corresponding author upon reasonable request.